# Title: Is Machine Learning Unsafe and Irresponsible in Social Sciences? Paradoxes and Reconsidering from Recidivism Prediction Tasks

**Authors:** Jianhong Liu*, Dianshi Li

**Affiliations:**

Faculty of Law, University of Macau; Macau, China

*Corresponding author. Email: jliu@um.edu.mo

**Abstract:**

The paper addresses some fundamental and hotly debated issues for high-stakes event predictions underpinning the computational approach to social sciences. We question several prevalent views against machine learning and outline a new paradigm that highlights the promises and promotes the infusion of computational methods and conventional social science approaches.

**Main Text:**

Recent advances and breakthroughs in artificial intelligence, such as machine learning (ML), particularly deep learning models, have shown their superior predictive performance and out-of-sample generalization capabilities (He et al., 2016; Vaswani et al., 2017). This progression has triggered social scientists to ponder the potential contributions of these technologies in their respective fields (Dwyer et al., 2018; Sun et al., 2020). Over the past decade, an array of applications has emerged, certain applications have blossomed into burgeoning disciplines, including computational psychology and computational education (Kučak et al., 2018; Rothenberg et al., 2023)

However, the deployment of these technologies has hit roadblocks in domains such as healthcare and criminal justice, where the opacity of machine learning models has raised concerns due to the high-risk nature of these domains themselves (Kirchner et al., 2016; Neri et al., 2020). Specifically, the public's trust to predictions in these high-stakes arenas—those with profound implications for individuals' or collectives' trajectories—are exceptionally susceptible to the ramifications of erroneous decision-making and obscured decision processes (Rudin et al., 2022; van Dijck, 2022), as erroneous judgments in these contexts might result in patients receiving inappropriate medical interventions or the inadvertent release of high-risk criminals, leading to potentially grave consequences (Garrett & Rudin, 2022; Rudin, 2019).

Despite considerable interdisciplinary efforts to integrate machine learning technologies into social science (Corbett-Davies et al., 2023; Simmler et al., 2022), recent scholarship has cast doubt on the very foundations of computational approach to social science. Researchers hailing from distinct academic traditions have grappled with finding a harmonious balance between data-driven methodologies and theory-driven frameworks (Angelino et al., 2017; Dwyer et al., 2018; Rothenberg et al., 2023; Sun et al., 2020). The focus of this multi-disciplinary discourse has been multi-faceted, encompassing the quest for algorithmic accuracy (Ozkan et al., 2020; Singh & Mohapatra, 2021), the imperative for interpretability in algorithmic decision-making (Rudin et al., 2022; Tolan et al., 2019), and the ethical dimensions related to procedural justice (Kaissis et al., 2020; Neri et al., 2020). Nevertheless, emerging literature has not only questioned the efficacy of these cross-disciplinary endeavors but also challenged the very underpinnings of the whole computational approaches to social science disciplines (Dressel &

Farid, 2018; Kirchner et al., 2016; Ozkan et al., 2020; Tolan et al., 2019; Wexler, 2017).

The views against machine learning approaches are primarily on two dimensions.

Firstly, the question of algorithmic accuracy remains a significant hurdle. A thorough systematic review reveals that most studies report only modest performance metrics. The Area Under the Receiver Operating Characteristic (AUC-ROC) scores typically hover around 0.74[1] (Travaini et al., 2022). Dressel & Farid (2018), using merely two features, achieved an accuracy of 68%, equivalent to the performance of an untrained human (Dressel & Farid, 2018). Similarly, Etzler et al. (2023) argue that the random forest—a quintessential black-box machine learning algorithm—doesn't surpass logistic regression in recidivism predictions. This skepticism is especially pertinent given that the promise of these algorithms largely hinges on their ability to offer robust predictive performance and generalize well to out-of-sample data (Dwyer et al., 2018; Herm et al., 2023).

Secondly, achieving acceptable levels of interpretability in algorithms remains a formidable challenge. For years, the scientific community has grappled with a trade-off between algorithmic performance and interpretability (Došilović et al., 2018; Gunning & Aha, 2019). Generally, results show that the more interpretable an algorithm is, the simpler—and often, the less accurate—it tends to be (Arrieta et al., 2020; Nanayakkara et al., 2018). The importance of interpretability or explainability in machine learning algorithms is a point of consensus across both computer science and social science literature (Minh et al., 2022; Skeem & Lowenkamp, 2020). This issue is particularly acute in high-stakes applications like recidivism prediction, where interpretability is seen as a crucial countermeasure against potential biases, including racial discrimination and gender inequality (Garrett & Rudin, 2022; Karimi-Haghighi & Castillo, 2021; Rudin et al., 2020a, 2020b; van Dijck, 2022). Consequently, algorithms that lack sufficient interpretability are likely to be deemed unsuitable for deployment in such sensitive and high-stakes domains (Rudin, 2019; Wang et al., 2022).

Given these formidable difficulties, however, this paper takes a distinctly different stance from much of the prevailing literature. By revisiting recent advancements in both AI and computational social science, this paper uncovers a series of paradoxes inherent in prior research. Grounded in these paradoxes, this paper contends that the commonly cited challenges — such as performance limitations, the black-box nature of algorithms, and issues of ethical requirements — are, in fact, significantly *overstated*. The quest for algorithmic interpretability, a concept almost revered by both computer and social scientists, should not be seen as the determining solution to address concerns of machine bias, potential discrimination, and algorithmic irresponsibility, especially when it comes to black-box algorithms. Instead, we advocate for a paradigm shift in the research. In this new paradigm, concerns over interpretability, accountability, and justice can be addressed holistically by integrating data-driven methods with theory-driven approaches. We argue that the current criticism of algorithms is mostly not appropriate. Solutions to these issues call for collaborative efforts of both computer scientists and social scientists.

---

[1] The systematic review indicates a reasonably good accuracy performance with an average score of 81%. However, it's challenging to label the classifier as "good" based solely on accuracy (ACC), as it's not an ideal metric for evaluating classification tasks. The ACC score can be significantly influenced by threshold variations without necessarily improving classification capability. A more robust metric is the area under the receiver operating characteristic (AUC-ROC) score. The score of 0.74 is not excellent for recidivism prediction, as it implies a heightened risk of inaccurate predictions.

## I. The Myths of Machine Learning's Weak Performance in Recidivism Prediction

*Paradox 1: Mistaking the Dated Algorithms for the Advanced*

A popular argument is that Algorithms do not perform well. However, an intriguing observation emerges upon scrutinizing the landscape of recidivism prediction studies: there is a pronounced reliance on classical machine learning algorithms and statistical methodologies, many of which were proposed several decades in the past. To elucidate, according to (Travaini et al., 2022), Logistic Regression (LR), Random Forests (RF), and Support Vector Machines (SVM) as the most prevalent algorithms in recidivism prediction, were introduced in 1944, 1984, and 1964 respectively (Berkson, 1944; Breiman et al., 1984; Vapnik, 1964). This reliance on dated algorithms raises questions about their contemporary relevance and performance.

Examining the trajectory of fields like computer vision and natural language processing provides a revealing contrast. These domains witnessed revolutionary progress upon the integration of deep learning techniques. Advanced architectures like ResNet in 2016 (He et al., 2016) and Transformer in 2017 (Vaswani et al., 2017) marked significant inflection points, ushering in unprecedented advancements in AI. These strides were made possible by innovative mechanisms like residual connections (He et al., 2016) and self-attention mechanism (Vaswani et al., 2017) facilitating the development of sophisticated models adept at challenges such as zero-shot learning and semantic segmentation (Guo et al., 2018; Wang et al., 2019).

Nevertheless, in the realm of recidivism prediction, there appears to be a hesitance to incorporate these cutting-edge approaches or technologies based on deep learning backbone, such as attention mechanism and transfer learning, which would significantly strengthen generalization ability out of sample (Vaswani et al., 2017; Zhou et al., 2022). The prevalent conclusions that algorithms failed to surpass human being and challenging entire efforts made in algorithmic recidivism prediction, solely based on dated classical algorithms, while overlooking the capabilities of contemporary techniques like deep learning, seems iniquitous. Especially when considering that even in their foundational domains, computer vision and natural language processing, classical algorithms like LR, RF, and SVM did not achieve commendable results until the recent paradigm shifts in deep learning.

### *Paradox 2: Feeding Machine Learning Algorithms by Data Collected from Traditional Scales and Questionnaires.*

Historically, in the evolution of criminal justice and legal research, scholars have predominantly employed tabular data for empirical analyses. Such data predominantly stems from official records, victimization surveys, and self-report questionnaires (Liu, 2008). Owing to the constraints of computational power and the desire to avoid "information overload", social scientists have traditionally narrowed their focus, often concentrating on a limited set of factors underpinned by one or a few analytical frameworks (Dwyer et al., 2018; Rothenberg et al., 2023). These practices align with a conventional paradigm that seeks to construct broad theoretical frameworks elucidating social phenomena or individual behaviors. Subsequently, these frameworks spawn empirically testable hypotheses that either corroborate or challenge the initial theories—a process emblematic of theory-driven methods (Bronfenbrenner & Morris, 2007; Sun et al., 2020). For instance, within Asian Criminology, scholars might posit that disruptive relations are significant predictors of criminal risk, according to relational approaches (Liu, 2017, 2021; Liu et al., 2017). This approach is intrinsically tied to the philosophical essence of theory-driven methods, which seek to distill generalizable patterns from empirical data.

In stark contrast to theory-driven methods, data-driven approaches are not burdened by the computational and analytic constraints inherent in more traditional statistical techniques. Consequently, data-driven machine learning predictors can encompass a vast array of potential influencing factors, pinpointing those with the most significant impact (Rothenberg et al., 2023). For instance, automatic content analysis, a machine learning-based data-driven method, is supplanting traditional content analysis due to its superior accuracy, cost-effectiveness, and expedited processing (Grimmer & Stewart, 2013; Wankhade et al., 2022). By extracting insights directly from raw content, rather than relying on coders or interviewers as intermediaries—as is customary in traditional content analysis and questionnaires—measurement biases, potential prejudices, and deviations arising from coders' limited perspectives are minimized (Drasgow, 1987; Zhao et al., 2022; Zhao et al., 2013). Algorithms excel at processing and modeling the vast range of unstructured data that would overwhelm human capacities (Wankhade et al., 2022). This divergence from the conventional theory-driven paradigm underlines the philosophical underpinnings of data-driven methods, which focus on intricate modeling processes to unveil the nuanced mechanisms concealed within empirical data (Dwyer et al., 2018; Rothenberg et al., 2023). Consequently, it's imperative for data-driven methodologies to be nourished with raw, comprehensive, and intricate data, rather than the data with handcraft feature typically harvested from psychological scales and surveys.

Unfortunately, reviewing data sources typically utilized for recidivism prediction reveals that a majority of such datasets can be categorized as outlined by Liu (2008), namely official records, victimization surveys, and self-reported questionnaires, which can be regarded as tabular datasets. Although some scholars regarding such problem, however, they just take it as the advantage of simple algorithms (usually more interpretable) in comparison to complex black-box algorithms (Garrett & Rudin, 2023), as simple algorithms perform as good as complex algorithms on simply tabular data (Garrett & Rudin, 2023; Herm et al., 2023). Such data predominantly emanate from scales and questionnaires, leading us to our paradox 2. However, in the realm of social science research, there seems to be no dearth of data suitable for training data-driven machine learning models. Owing to specific historical, cultural contexts, and scientific training, criminology scholars in Asia exhibit a predilection for interview-centric qualitative methodologies as one of the primary methodologies in criminal justice and criminology research—a methodology once deemed antiquated (Zhang & Liu, 2023). Fortunately, with the advent of unstructured textual and audiovisual AI analytical tools, these interview archives can be transformed into invaluable resources for data-driven machine learning investigations, aiding in the measurement of deeper psychological facets, such as personality traits and charisma (Harrison et al., 2019, 2020), which offer more information for analysis. Furthermore, with the emergence of smart prison systems and judicial frameworks within smart city infrastructures (Simmler et al., 2022), the profusion of unstructured data stands poised to offer a holistic assessment of an offender's rehabilitation during incarceration and the associated risks post-release (Korobkin & Ulen, 2000).

*Paradox 3: Historical Factors Determines Future Human Behavior*

Another potential constraint on the efficacy of prediction pertains to the foundational behavioral assumptions underpinning the task. In recent works, most of the scholars, specially who with computer background, trained recidivism prediction algorithms by utilizing historical factors (like age, historical criminal records) and the binary outcomes of recidivism behavior in a fixed period of time (like 6 months or 2 years) (Dressel & Farid, 2018; Etzler et al., 2023; Garrett & Rudin, 2022; Ma et al., 2022; Ozkan et al., 2020; Rudin et al., 2022; Wang et al., 2022). Initially, those scholars employ these historical elements to forecast whether the criminal would re-offend. Subsequently, the binary outcome of recidivism serves as a proxy variable for recidivism risk. Some computer scientists also employ the probability (or score) assigned by the model for an offender's likelihood of re-offense as a gauge for their recidivism risk (Etzler et al., 2023; Ma et al., 2022; Wang et al., 2022). While such configurations may seem intuitively compelling, they often embody an oversimplified and deterministic viewpoint, which stands in contradiction to contemporary social science theories.

Firstly, historical factors alone are insufficient predictors of human actions. This notion harks back to the rationality assumption prevalent in classical economics, which sought a formulaic representation of human behavior prediction and simulations (Korobkin & Ulen, 2000). Such an approach is now deemed an oversimplification of actual socio-economic dynamics. Modern behavior science theories underscore the uncertainty and bounded rationality of decision making, for instance, anchoring, heuristics, framing, and other effects suggests that oversimplifying behavioral assumption is much harmful to accurately describing human behavior (Kahneman & Tversky, 1972; Tversky & Kahneman, 1981).

Social sciences provide ample evidence that only historical factors alone is not able to predict binary outcomes of behavior without additional contextual information and prediction can only be probabilistic, regardless of which advanced algorithm is employed. Whether an individual re-offends post-release is influenced by myriad factors including personality, background, environment, education, opportunities, social networks, mental health, and unforeseen events (Liu, 2017, 2021; Messner et al., 2018). These future-oriented determinants, distinct from historical data, also play a crucial role in influencing re-offense.

In designing predictive models, it is imperative to acknowledge the unpredictability and uncertainty inherent in human behavior concerning both predictive targets and elements. Integrating prior knowledge from social sciences, both in terms of model structure and optimization objectives, can align the algorithm's assumptions more closely with the nuances of real-world socio-systems. It's also vital to recognize that the risk of recidivism evolves with time, environment, and unforeseen factors. Hence, the risk should not be "Bernoullized" into a binary variable, nor should it be used as a definitive label for offenders (Wang et al., 2022).

## II. Worship of Interpretability of Algorithms

*Paradox 4: Interpretability of Algorithms Produce Justice, Fairness, Non-discrimination, Reliability, Trust by the Public, and Other Ethical Requirements.*

Almost all the research articles assert the importance and necessity of interpretability of ML algorithms, especially when it comes to its application in social science (Li et al., 2019; Medvedeva et al., 2019; Padovan et al., 2023; Travaini et al., 2022; van Dijck, 2022). An influential publication by ProPublica (Kirchner et al., 2016) accused that algorithms discriminate against the black. ProPublica is an NGO media, which focuses on reporting abuse of power and inequality. The publication created a backlash against AI (Chodosh, 2018; Johndrow & Lum, 2019; Wexler, 2017). Rudin et al. cite ProPublica's case and indicate that, although there are statistical flaws in ProPublica's results, it is hard for ML algorithms to gain public trust unless more interpretable and fair models are proposed (Rudin et al., 2020a, 2020b). Since then, an increasing number of scholars are asserting the importance of interpretability; it seems become widely accepted that interpretable algorithms provide solutions to the issue of racial and gender discrimination (Johndrow & Lum, 2019; Soares & Angelov, 2019), unfairness (Etzler et al., 2023; Johndrow & Lum, 2019; Rudin et al., 2020a), distrust from public (Rudin et al., 2022; Wang et al., 2022), the need of clarify rights and responsibilities (Rudin et al., 2020b), promoting reliability and validity (Wang et al., 2022), and even enhance the generalization ability on data with different distributions (Rudin et al., 2022).

However, we argue that interpretability's concept (Arrieta et al., 2020; Marcinkevics & Vogt, 2020), functions (Herm et al., 2023; Rudin, 2019), explanation typologies (Mohseni et al., 2021), and even the probability that it can be achieved (Bathaee, 2017; Castelvecchi, 2016), are ambiguous. Although the transparency of algorithms is intuitively attractive (Ghassemi et al., 2021), it might not be very sound to claim that algorithmic interpretability is the definitive answer to the issues raised.

Nearly all literature implicitly assumes that explainability or interpretability can mitigate potential issues of inequity, injustice, discrimination, and public distrust embedded within algorithms (Medvedeva et al., 2023; Rudin, 2019; Rudin et al., 2020a; Wang et al., 2022). Some influential authors claim utmost importance of interpretability:

*To prevent errors, prevent due process violations, allow independent validation of models, and gain public trust, we **must** create interpretable and fair models."(Wang et al., 2022)*

*"This further complicates their adoption in many sensitive disciplines, raising concerns from the ethical, privacy, fairness, and transparency perspectives. The **root cause** of the problem is the lack of explainability and/or interpretability of the decision."(Islam et al., 2020)*

In recent works, several studies even posit explainability or interpretability as the sole solution to achieving these aims (Islam et al., 2020; Padovan et al., 2023; Rudin et al., 2020a). Based on these claims, researchers have proposed so-called algorithms for detecting discrimination, metrics for assessing fairness, and indiscriminate behavior prediction models (Green, 2020; Karimi-Haghighi & Castillo, 2021; Wang et al., 2022). Few scholars challenge this widely accepted, almost consensual, premise.

An simple example is from recent work by Rudin's group which presents a paradoxical challenges such assumptions (Wang et al., 2022). In their study, Wang et al. employed the Explainable Boosting Machine (EBM) models to predict recidivism, asserting that their models are interpretable, fair, and potentially valuable in real-world applications. Their primary approach to elucidate the model's results was through visualization, and by enumerating the importance of each factor influencing recidivism predictions. Intriguingly, the variable "ADE" (which represents the number of times an individual was assigned to alcohol and drug education classes—mandated by Kentucky state law for any DUI conviction but doesn't necessarily indicate successful completion) was deemed more influential than "p_drug" (indicative of prior drug-related charges). Nevertheless, we believe that any experienced social science scholars would surely consider that "p_drug" signifies a more severe offense than "ADE". Such outcomes, devoid of theoretical grounding and logical interpretation risk muddling public perception rather than fostering understanding.

Another point of contention revolves around the interpretation of logistic regression. Owing to its straightforward structure, logistic regression is often categorized as a white-box model (Freitas, 2019; Zhou et al., 2018). All pertinent information is encapsulated within the coefficients of logistic regression models, and the computational process of the algorithm is wholly transparent. But it's overly simplistic to claim that logistic regression inherently ensures fairness, justice, and other ethical requirements. Recent studies have raised alarms about the potential pitfalls of logistic regression in high-stakes scenarios. Notably, its coefficients can be disproportionately influenced by imbalanced datasets and sensitive attributes, potentially compromising its fairness (Alikhademi et al., 2021; Dreiseitl & Ohno-Machado, 2002). Even with perfect performance, the fairness of algorithms can be compromised by potential biases lurking in the data. Completely interpretable algorithms alone are insufficient to address the inherent inequalities in the real world, unless someone actively works to advance social structures, rather than merely addressing unbalanced data in the lab.

These two examples serve as a reminder that algorithmic outputs, even from models touted as "interpretable", need to be contextualized and critically examined. There is a huge chasm between interpretable models and the ethical hopes. While the transparency of algorithms largely equates to technical interpretability (Arrieta et al., 2020; Stevens & De Smedt, 2023), it remains at a distinct remove from a genuine human-centric understanding. And even more profound is the gap between this human-centric understanding and the broader, yet crucial, benchmarks of justice, fairness, non-discrimination, reliability, and public trust. Simply achieving transparency is insufficient, and even pairing it with understanding doesn't necessarily meet the higher ethical and societal standards demanded by the public. Regrettably, a majority of scholars in this domain, particularly those with backgrounds in computer science, often overlook this distinction. They tend to view technical interpretability as the end and target of their research, and mistakenly equate explainability/interpretability or understandability with ethical benchmarks. Their focus on interpretability is predominantly anchored in elucidating models or datasets, rather than unpacking the broader societal phenomena. This oversight is elaborated upon in Paradox 5:

*Paradox 5: It Is Enough to Explain or Interpret Algorithm.*

Computer Scientists and social scientists have been considered that explanations or interpretations being faithful to the original model computed is important in recent works (Rudin, 2019; Rudin et al., 2020a). However, scholars seem to have missed a crucial distinction: the explanations or interpretations of algorithms do not necessarily equate to the explanations or interpretations of social phenomena. Meeting the ethical requirements of algorithms largely depends on the explanations or interpretations of social phenomena rather than on the algorithms themselves.

The statement "All 500 individuals with a credit history of less than 5 years were predicted to default on a loan" is taken from an article by Rudin (Garrett & Rudin, 2023). Rudin, a leading advocate for algorithm interpretability, introduced a novel inherently interpretable algorithm in her article. This algorithm is said to generate faithful explanations, touted to be more useful and safer than black-box models. However, her example of 500 default on loan did not really provide a good example for her purpose, and it falls short of answering such a crucial question: Why are these individuals deemed likely to default on loans? This is akin to noting that 500 patients recovered after taking a mysterious medicine. We remain unaware of the medicine's workings or its interaction with the ailment; we only know that many have recovered post-administration.

Algorithms or models of this nature are indeed inherently interpretable. Their interpretations remain largely faithful to the model and even to the raw data. However, they essentially function as a new black box, failing to address the underlying "why" behind their conclusions. To some extent, even the interpretation of the algorithm remains to contain further black box, as no explanation can completely elucidate all mechanisms in human inquiry. We might term these algorithms "black-box interpretability algorithms".

Black-box Interpretability Algorithms (BIA) are nearly as unsatisfactory as the original black-box algorithms, while their interpretations may seem meaningful to some extent, they do not resolve the essential problem of black box. Firstly, BIA struggles to address undesirable biases present in the training data. While numerous so-called fairness metrics or benchmarks have attempted to assert their roles in combating inequality (Ma et al., 2022), a systematic analysis has revealed two primary fairness categories: (1) those that limit the decision effects on disparities and (2) those that restrict the decision effects based on legally protected characteristics, such as race and gender. Alarmingly, these two fairness approaches can inadvertently harm the very groups they aim to shield (Corbett-Davies et al., 2023).

Secondly, BIA blurs the distinction between interpretations related to social phenomena, the real world, and the algorithms themselves. Rudin frequently references the Rashomon effect in machine learning, suggesting that within a group of nearly identical-performing models, there might exist at least one simpler (and perhaps more interpretable) model. While these models may showcase similar performance metrics, they can offer vastly different interpretative narratives (Breiman, 2001; Müller et al., 2023; Rudin, 2019). Yet, if we adhere to the notion that a singular truth exists, the emergence of multiple, seemingly aligned interpretations introduces inherent weaknesses. Interestingly, scholars studying the Rashomon effect posit that none of the multiple interpretations of a single dataset can be confidently declared as the definitive truth (Rudin et al., 2022). Thus, how can we expect interpretability, which may not align with truth, to enlighten us on fairness, justice, and other ethical considerations?

More critically, the pursuit of BIA obscures alternative approaches to meeting ethical standards. Proposing a black-box algorithm might seem to invite criticism and be perceived as irresponsible. Yet, both BIA and black-box algorithms similarly fall short in addressing ethical requirements, potentially resulting in increased operational costs for institutions and diminished engagement with non-interpretable technologies (Bathaee, 2017). Genuine ethical-based interpretability algorithms should guide scholars in understanding social phenomena and delving into the intricate mechanisms of the real world, rather than merely focusing on algorithmic rules. In essence, explainable or interpretable algorithms should offer insights that transcend the algorithms themselves, a notion that leads us to propose Paradox 6.

### *Paradox 6: Interpretability for Computer Scientists and Engineer Is Same as that for the Social Scientists, the Authorities, and the Public.*

Most systematic reviews converge on the idea that true explainability hinges on providing human-centric, lucid explanations, and emphasize the primacy of the end users—the human audience—over mere structural simplicity of the models (Adadi & Berrada, 2018; Arrieta et al., 2020; Islam et al., 2020; Ras et al., 2018; van der Waa et al., 2021; Zhou et al., 2018). However, a fundamental oversight in these deliberations is the question: Who exactly is the user? Social scientists? Regulatory authorities? The general public? Oftentimes, discussions inadvertently conflate these diverse entities, resulting in an oversimplified dialogue that fosters misaligned expectations between algorithm developers and the broader audience (Mohseni et al., 2021). The dilemma is that, given the plethora of definitions of explainability stemming from diverse algorithmic origins, the interpretative significance varies for computer scientists, social scientists, legal professionals, governmental entities, and the general public. Which form of explainability should we then adopt, while it appears that not all kinds of interpretations are considered useful (Ghassemi et al., 2021; Mohseni et al., 2021) .

For analytical convenience, as a simple example, we categorize users into four groups: computer scientists, social scientists, regulatory authorities, and the public. From a prior knowledge standpoint, computer scientists possess deep technical expertise but may lack nuanced understanding of sociolegal constructs like fairness, justice, and trust. Social scientists, conversely, maybe less technologically versed but hold profound insights into these socio-cultural dimensions. Regulatory authorities, being the ultimate end-users of XAI, have practical considerations extending beyond mere academic theories (Ghassemi et al., 2021; Mohseni et al., 2021). Lastly, the general public, the end beneficiaries of these models, often lack both technical expertise and specialized sociological training, focusing primarily on their rights and obligations (Miller, 2019; Miller et al., 2017).

Despite the proliferation of explainability techniques, from post-hoc rationales to visualizations and local explanations (Arrieta et al., 2020; Li et al., 2022; Stepin et al., 2021), no single method can satiate the diverse palate of all user groups (Ghassemi et al., 2021; Mohseni et al., 2021). Even universally recognized, transparent models like logistic or linear regression—extensively used across social sciences—are comprehendible predominantly to those with academic training (Brożek et al., 2023). A college undergraduate without mathematical, statistical, or social science training might struggle to interpret such models, let alone the layperson (Batanero et al., 1997; Engel & Sedlmeier, 2011; Zhao & Zhang, 2014).

In essence, each user group's demands from interpretability are distinct (Ghassemi et al., 2021; Mohseni et al., 2021). Computer scientists seek explanations to enhance performance and rectify flaws (Amann et al., 2020). Social scientists aim to measure constructs like fairness and justice, abstracting broader social principles from data-driven models. Regulatory bodies focus on accountability, a domain where legal scholars play a critical role. The public seeks assurance on their rights, essentially translating to trust in regulatory bodies. Expecting the general populace to grasp the intricacies of AI and XAI is an unrealistic aspiration (Miller, 2019; Miller et al., 2017).

From this perspective, whether the public trusts algorithms fundamentally depend on their confidence in regulatory institutions, which fall within the academic realms of communication and public administration. Authorities should major and elucidate the performance, fairness, and justice of these models, fostering public trust. Instead of hoping for simpler models at the expense of performance, the ideal model, characterized by enhanced performance, fairness, and non-discrimination, should be a collaborative endeavor between computer and social scientists (Miller, 2019).

*Paradox 7: Interpretability Is the Sole Way that Leads to Fairness, Justice, Non-discrimination, and Other Ethical Requirements.*

In contemporary scholarly discourse, a proliferation of literature has emerged advocating for the deployment of interpretable models as a panacea for achieving fairness, justice, and non-discrimination (Jo et al., 2023; Ma et al., 2022; Rudin, 2019; Rudin et al., 2022; Wang et al., 2022). The conflation of interpretability with these lofty ideals is not merely prevalent but is often portrayed as their sole arbiter. However, there exists scant empirical evidence substantiating the claim that interpretability invariably culminates in these desired outcomes (Brożek et al., 2023; Ghassemi et al., 2021; Herm et al., 2023; Miller et al., 2017).

A more circumspect perspective posits that the quest for interpretability invariably entails simplifying or summarizing the behaviors of complex models to render them palatable to human comprehension (Garrett & Rudin, 2022; Wang et al., 2022). This introduces an intricate trade-off: the aspiration to distill models for heightened interpretability might be at loggerheads with the imperative to preserve their predictive prowess (Jo et al., 2023; Minh et al., 2022). Amplifying a model's predictive accuracy often necessitates a commensurate augmentation of its complexity. Human societies, as intricate amalgamations of economic, technological, political, legal, social, historical, cultural, geographical, climatic, and ecological facets, ensure human behaviors are invariably influenced by a mélange of anticipated and unanticipated, deterministic and stochastic variables, exhibiting non-linear and time-variant characteristics(Hong & Wang, 2021; HONG & WANG, 2023). Consequently, hyper-dimensional, hyper-parametric, and profoundly intricate models become indispensable for robust predictions, particularly for out-of-sample forecasting (Dwyer et al., 2018; Yongmiao & Shouyang, 2021).

Given humanity's inherent incapacity to fathom raw data or the labyrinthine intricacies of economic and social machinations, resorting to simplification and abstraction in theoretical constructs, how then can one harbor realistic expectations of representing the extremely complex machinations with a glass-box or white-box model? And how then can one harbor realistic expectations of completely explicating such a model?

While the primary thrust of this article is not to wade into the philosophical debates surrounding the feasibility of AI interpretability, it seeks to recalibrate the misplaced expectations that humanity can solely leverage AI's interpretability techniques to instantiate fairness, justice, trustworthiness, and accountability.

In essence, humanity's tryst with "black boxes" is neither novel nor unprecedented, especially when venturing into high-stakes domains.

Throughout scientific history, numerous phenomena were harnessed and applied long before their complete comprehension materialized. A pertinent analogy in the realm of computational social science, or more specifically in the context of recidivism prediction, is the use of medical drugs and devices (Ghassemi et al., 2021). Acetaminophen, known for its analgesic properties, has been a staple in the medical community for over a century. Amid the recent onslaught of COVID-19, acetaminophen retained its stature as one of the most efficacious and safest remedies for alleviating fever symptoms (Galluzzo et al., 2023; Yousefifard et al., 2020). Intriguingly, the mechanistic underpinnings of acetaminophen remain enigmatic, an embodiment of the "black box" paradigm (Kirkpatrick, 2005; Kis et al., 2005). Yet, this hasn't deterred regulators, physicians, or patients from placing unwavering trust in its safety profile.

The medical community, in its endeavor to circumnavigate the intrinsic complexities of these "black boxes" and instill trust among regulators, practitioners, and patients, has resorted to the rigorous scientific methodology of Randomized Controlled Trials (RCTs) (Ghassemi et al., 2021). RCTs assess the efficacy of interventions, be they novel pharmaceuticals, therapeutic regimens, or preventative strategies. Participants are judiciously randomized across experimental groups (subjected to the intervention) and control groups (either receiving a placebo or the standard treatment). Such randomization ensures homogeneity at baseline, thereby mitigating biases. The juxtaposition of the control group allows for an empirical comparison of the intervention's efficacy against either the absence of intervention or alternative interventions. Many RCTs incorporate a "double-blind" design, wherein neither the participants nor the investigators are privy to group allocations, a stratagem that considerably attenuates biases and subjectivity. Upon trial culmination, investigators juxtapose outcomes across groups to infer the intervention's efficacy (Concato et al., 2000; Deaton & Cartwright, 2018). Historically, RCTs have been heralded as the "gold standard" in clinical trials, their design meticulously calibrated to minimize biases and errors, thereby proffering robust empirical evidence even when the intervention remains a "black box" (Bothwell et al., 2016; Torgerson et al., 2015).

Fortuitously, the realm of social sciences is replete with analogous methodological tools. An array of counterfactual-based causal inference tools endows us with the capacity to assess policy effects rigorously. More precise post-hoc inferences fortify our ability to evaluate algorithmic fairness, both from input and output perspectives. Emerging techniques, such as simulations grounded in authentic datasets, show promise in becoming the de facto "gold standard" for validating predictive tasks within social sciences (Zhao et al., 2022). Drawing upon research in these avant-garde domains, there's a compelling rationale to posit that, analogous to the medical sector, the "black box" of AI in social sciences can also be underpinned by trust and accountability mechanisms. And the first step of this is to acknowledge the existing of BIA, and recognize it is not better than completely black box. The mode of RCTs and post-hoc inferences methods are still a kind of powerful and another tool distinguished from XAI for helping scholars and individuals exploring the unknown knowledge and world. The realization of this vision mandates a collaborative odyssey, uniting computer scientists and social scientists alike, paving alternative avenues—distinct from sheer interpretability—to ensure fairness, trustworthiness, accountability, and other ethical requirements of AI models.

## III. Conclusion

This paper critically examines the application of machine learning algorithms in the field of social sciences, particularly focusing on their role in high-risk decision-making. We challenge prevalent views on performance, interpretability, and accountability, believing that the complexities of these issues might be overly magnified. The aim of this study is to offer a fresh perspective on integrating machine learning with social science research. We recognize that these critical viewpoints require further empirical evidence for validation. To this end, we intend to undertake in-depth empirical research in our upcoming series of articles.